\documentclass[twocolumn]{aastex62}

\graphicspath{{./}{figures/}}

\received{January 1, 2018}
\revised{January 7, 2018}
\accepted{\today}
\submitjournal{ApJL}

\shorttitle{2019 flare in OJ~287}
\shortauthors{Laine et al.}

\begin{document}

\title{
Spitzer Observations of the Predicted Eddington Flare from Blazar OJ~287
}

\correspondingauthor{Lankeswar Dey}
\email{lankeswar.dey@tifr.res.in}

\author[0000-0003-1250-8314]{Seppo Laine}
\affiliation{IPAC, Mail Code 314-6, Caltech, 1200 E. California Blvd., Pasadena, CA 91125, USA}

\author[0000-0002-2554-0674]{Lankeswar Dey}
\affiliation{Department of Astronomy and Astrophysics, Tata Institute of Fundamental Research, Mumbai 400005, India}

\author[0000-0001-8580-8874]{Mauri Valtonen}
\affiliation{Finnish Centre for Astronomy with ESO, University of Turku, FI-20014 Turku, Finland}

\author[0000-0003-4274-4369]{A. Gopakumar}
\affiliation{Department of Astronomy and Astrophysics, Tata Institute of Fundamental Research, Mumbai 400005, India}

\author[0000-0003-3609-382X]{Stanislaw Zola}
\affiliation{Astronomical Observatory, Jagiellonian University, ul. Orla 171, Cracow PL-30-244, Poland}
\affiliation{Mt. Suhora Astronomical Observatory, Pedagogical University, ul. Podchorazych 2, PL30-084 Cracow, Poland}

\author{S. Komossa}
 \affiliation{Max-Planck-Institut f{\"u}r Radioastronomie, Auf dem H{\"u}gel 69, 53121 Bonn, Germany} 

\author{Mark Kidger}
\affiliation{PLATO Science Operations Centre, ESAC, European Space Agency, E-28691 Villanueva de la Cañada, Madrid, Spain}

\author[0000-0003-1758-1908]{Pauli Pihajoki}
\affiliation{Department of Physics, University of Helsinki, Gustaf Hällströmin katu 2a, FI-00560, Helsinki, Finland}

\author[0000-0003-4190-7613]{Jos\'e L. G\'omez}
\affiliation{Instituto de Astrof\'{\i}sica de Andaluc\'{\i}a-CSIC, Glorieta de la Astronom\'{\i}a s/n, 18008 Granada, Spain}

\author{Daniel Caton}         
\affiliation{Dark Sky Observatory, Dept. of Physics and Astronomy, Appalachian State University, Boone, NC 28608, USA}

\author[0000-0002-0712-2479]{Stefano Ciprini} 
\affiliation{Space Science Data Center—Agenzia Spaziale Italiana, via del Politecnico, snc, I-00133, Roma, Italy}
\affiliation{Instituto Nazionale di Fisica Nucleare, Sezione di Perugia, Perugia I-06123, Italy}

\author{Marek Drozdz}
\affiliation{Pedagogical University of Cracow, Mt. Suhora Astronomical Observatory, ul. Podchorazych 2, PL30-084 Cracow, Poland}

\author[0000-0002-8855-3923]{Kosmas Gazeas}
 \affiliation{Department of Astrophysics, Astronomy and Mechanics, National and Kapodistrian University of Athens, Zografos GR-15784, Athens, Greece}

\author[0000-0001-7668-7994]{Vira Godunova} 
\affiliation{ICAMER Observatory of  NASU, 27 Acad. Zabolotnoho Str., 03143 Kyiv, Ukraine}

\author{Shirin Haque}
\affiliation{Department of Physics, University of the West Indies, St. Augustine, Trinidad, West Indies}

\author{Felix Hildebrandt}
\affiliation{Astrophysikalisches Institut und Universit\"ats-Sternwarte, Schillerg\"a{\ss}chen 2, D-07745 Jena, Germany}

\author[0000-0002-7273-7349]{Rene Hudec}         
\affiliation{Czech Technical University in Prague, Faculty of Electrical Engineering, Technicka 2, Prague 166 27, Czech Republic}
\affiliation{27 Kazan Federal University, Kremlyovskaya 18, 420000 Kazan, Russian Federation}

\author{Helen Jermak}
\affiliation{Astrophysics Research Institute, Liverpool John Moores University, IC2, Liverpool Science Park, Brownlow Hill, L3 5RF, UK}

\author[0000-0002-5105-344X]{Albert K.H. Kong}
\affiliation{Institute of Astronomy, National Tsing Hua University, Hsinchu 30013, Taiwan}

\author{Harry Lehto }
\affiliation{Department of Physics and Astronomy, University of Turku, Finland}

\author[0000-0002-0490-1469]{Alexios Liakos}
\affiliation{Institute for Astronomy, Astrophysics, Space Applications and Remote Sensing, National Observatory of Athens, Metaxa and Vas. Pavlou St., Penteli, Athens GR-15236, Greece}

\author{Katsura Matsumoto} 
\affiliation{Astronomical Institute, Osaka Kyoiku University, 4-698 Asahigaoka, Kashiwara, Osaka 582-8582, Japan}

\author{Markus Mugrauer}
\affiliation{Astrophysikalisches Institut und Universit\"ats-Sternwarte, Schillerg\"a{\ss}chen 2, D-07745 Jena, Germany}

\author{Tapio Pursimo}
\affiliation{Nordic Optical Telescope, Apartado 474, E 38700, Santa Cruz de la Palma, Spain}

\author{Daniel E. Reichart}
\affiliation{University of North Carolina at Chapel Hill, Chapel Hill, North Carolina NC 27599, USA}

\author[0000-0003-0404-5559]{Andrii Simon}         
\affiliation{Astronomy and Space Physics Department, Taras Shevchenko National University of Kyiv, 64/13, Volodymyrska St.,  Kyiv, 01601 Ukraine}

\author{Michal Siwak}
\affiliation{Pedagogical University of Cracow, Mt. Suhora Astronomical Observatory, ul. Podchorazych 2, PL30-084 Cracow, Poland}

\author{Eda Sonbas}
\affiliation{University of Adiyaman, Department of Physics, 02040 Adiyaman, Turkey}




\begin{abstract}
Binary black hole (BH) central engine description for the unique blazar OJ~287 predicted that the next  secondary BH impact-induced bremsstrahlung flare should peak on 2019 July 31.
This prediction was based on detailed general relativistic modeling of the secondary BH trajectory around the primary BH and its accretion disk.
The expected flare was termed the {\it Eddington flare} to commemorate the centennial celebrations of now-famous solar eclipse observations to test general relativity by Sir Arthur Eddington.
We analyze the multi-epoch \texttt{Spitzer} observations of the expected flare between 2019 July 31 and 2019 September 6, as well as baseline observations during 2019 February--March. 
Observed \texttt{Spitzer} flux density variations during the predicted outburst time display a strong similarity with the observed optical pericenter flare from OJ~287 during 2007 September.
The predicted flare appears comparable to the 2007 flare after subtracting the expected higher base-level \texttt{Spitzer} flux densities at 3.55 and 4.49 $\mu$m compared to the optical $R$-band.
Comparing the 2019 and 2007 outburst lightcurves and the previously calculated predictions, we find that the {\it Eddington flare} arrived within 4 hours of the predicted time. 
Our \texttt{Spitzer} observations are well consistent with the presence of a nano-Hertz gravitational wave emitting spinning massive binary BH that inspirals along a general relativistic eccentric orbit in OJ~287. 
These multi-epoch \texttt{Spitzer} observations provide a parametric constraint on the celebrated BH no-hair theorem.
\end{abstract}



\section{Introduction} \label{sec:intro}

The International Pulsar Timing Array (IPTA) consortium aims to inaugurate the era of nano-Hertz (Hz) gravitational wave (GW) astronomy during the next decade \citep{perera2019}.
This is expected to augment the already established hecto-Hz GW astronomy by the LIGO--Virgo collaboration \citep{lvc2019} and the milli-Hz GW astronomy to be established by space-based observatories in the 2030s \citep{lisa2019}.
Massive black hole (BH) binaries, emitting nano-Hz GWs, are the most prominent IPTA sources \citep{sarah2019}.
Therefore, observational evidence for the existence of such binaries has important IPTA implications \citep{goulding2019}.

The binary black hole (BBH) central engine description for the bright blazar OJ~287 provides the most promising scenario for the existence of a nano-Hz GW emitting massive BH binary \citep{dey2019}.
The model naturally explains the observed double-peaked high brightness flares (outbursts) from OJ~287 and predicts the arrival time of future outbursts.
These flares arise due to the impact of an orbiting secondary BH onto the accretion disk of the primary.
In the resulting thermal flares, flux densities (hereafter ``flux")  in the UV--infrared wavelengths increase sharply within just a day or so and then fall off more slowly in the following days \citep{valtonen2019}.
Accurate timing of these flares allows us to track the general relativistic  trajectory of the secondary BH 
and to determine BBH central engine parameters \citep[][hereafter D18]{dey2018}.

The nature of such flares and the method of predicting future flares were detailed by \citet{lehto1996} and \citet{sund1997}.
In their model, the secondary BH plunges through the accretion disk twice per orbit, which ensures two flares per period.
This model also predicted that impact flares should be thermal, with a flat bremsstrahlung spectrum, rather than the ubiquitous synchrotron flares with a power-law spectrum.
It was not a trivial prediction, as no bremsstrahlung flares had been observed in any blazar up to that time.
The observations of the 2005 November flare confirmed this prediction \citep{valtonen2006,valtonen2008a,valtonen2012}. 
This was followed by a successful observational campaign, launched to monitor the predicted pericenter flare of 2007 \citep{valtonen2008b}.
These observations demonstrated the importance of incorporating the effects of quadrupolar order GW emission while predicting the impact flare epochs from the blazar.
Further, the successful observation of OJ~287's 2015 apocenter impact flare, predicted by \citet{valtonen2011}, provided an estimate for the spin of its primary BH \citep{valtonen2016}.
The present BBH model, extracted from the accurate timing of 10 flares between 1913 and 2015 (D18), is specified by the following parameters:
primary with mass $1.835\times10^{10}\,M_{\odot}$ and Kerr parameter $a=0.38$, and a $1.5\times10^{8}\,M_{\odot}$ secondary in an eccentric ($e\sim0.65$) orbit with a redshifted orbital period of $12$ years.

D18 predicted that the next impact flare from OJ~287 should peak in the early hours of 2019 July 31, UT, within a specified time interval of $\pm4.4$ hours ({\it Eddington flare}).
This prediction is fairly unique as there are no free parameters whose value can be constrained from the actual observations of the flare, in contrast to the earlier flares.
Ideally, we would have launched a ground-based optical observational campaign to monitor the predicted {\it Eddington flare}.
However,  OJ~287 was at a solar elongation $<5^\circ$ during 
the peak of the flare. Therefore, there was no option to confirm it by means of a ground-based observing campaign. 
The \texttt{Spitzer} space telescope, operating at infrared wavelengths, turned out to be the best substitute for optical monitoring.
An earlier optical/infrared campaign was organized for flux normalization.
In what follows, we explain why we are confident about the presence of the predicted {\it Eddington flare} in our \texttt{Spitzer} data and state its implications.

\section{BBH central engine of OJ~287 and its 2019 Prediction} \label{sec:BBH_model}

The 130 years long optical lightcurve of OJ~287 reveals two prominent outbursts every 12 years \citep{dey2019}. 
The outburst timings are consistent with a scenario in which bi-orbital secondary BH impacts generate two hot bubbles of plasma on either side of the primary BH accretion disk.
These bubbles expand and eventually become optically thin. At this epoch, the radiation from the entire bubble volume is released and we  observe a big thermal flare.
In the model, the observed  steeply rising flux during a flare arises from an increase in the visible radiating volume, while the declining flux comes with the decreasing temperature from the associated adiabatic expansion.
Both processes should produce radiation that is wavelength independent while  timing various epochs of the flare.

\begin{figure}
    \centering
    \includegraphics[scale = 0.32]{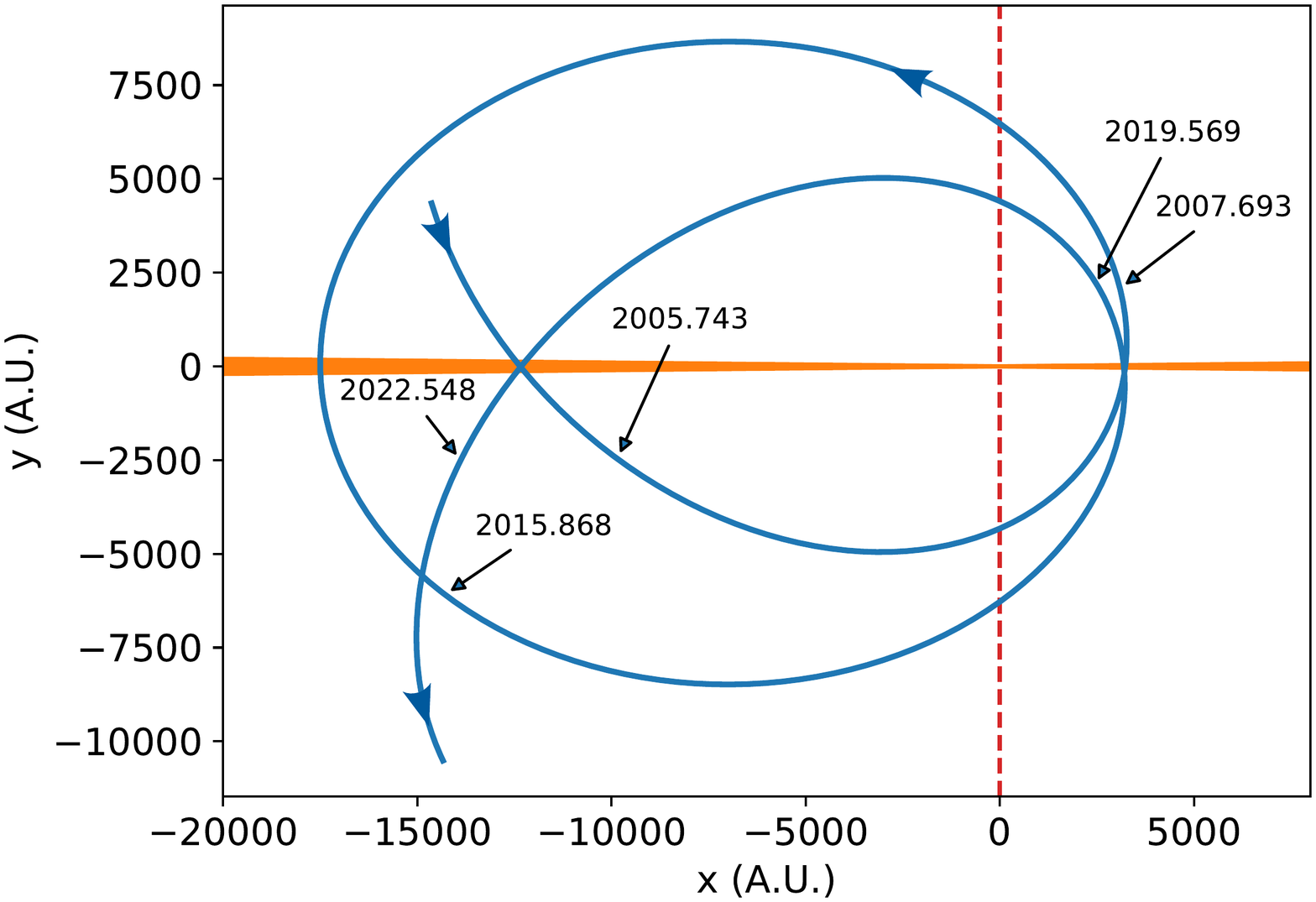}
    \caption{ General relativistic  orbit of the secondary BH in OJ~287 during the 2005–2023 window (D18). The primary BH is situated at the origin with its accretion disk in the $y=0$ plane.
    The impacts that caused the 2007 and 2019 outbursts happen to originate roughly from the same location of the disk near the pericenter, and the secondary BH follows similar trajectories, leading to fairly identical lightcurves.
    In contrast, the 2005 and 2022 impact flare lightcurves are expected to be different.
    The orbit is calculated using our PN accurate binary BH description.}
    \label{fig:orbit}
\end{figure}

In general, the points of impact are located at different distances from the primary due to the general relativity (GR) induced pericenter advance \citep{lehto1996}.
However, there are occasions during which two impacts happen close to the pericenter of such a relativistic orbit.
We expect that the astrophysical conditions are fairly similar at such impacts, leading to essentially similar flares.
The orbit solution of D18 shows a pair of pericenter flares during 2007 September 14 and 2019 July 31 (Figure~\ref{fig:orbit}).
This allowed us to use the observed optical lightcurve of the 2007 outburst as a template to analyze our \texttt{Spitzer} observations of OJ~287 during late July and early August 2019.

A post-Newtonian(PN) approximation to GR is employed 
to track the secondary BH orbit around the primary BH \citep{WM2017}.
We incorporate higher-order corrections to both the conservative and reactive contributions to the relative acceleration ${\bf\ddot{x}}$ (see Equation~1 in D18). 
Crucially, these corrections involve certain GW emission induced  ${\bf\ddot{x}}_{\rm4PN\,(tail)}$ contributions due to the scattering of quadrupolar GWs from the space-time curvature created by the total mass (monopole) of the system \citep[D18, ][]{BS_93}.
Additionally, we incorporate various spin induced contributions to ${\bf\ddot{x}}$ that arise from general relativistic spin-orbit and the classical spin-orbit interactions. 
The latter contributions depend on the quadrupole moment of the primary BH and affect the expected outburst time of the {\it Eddington flare}.
Therefore, the accurate determination of the epoch of this flare has the potential to constrain the celebrated  BH no-hair theorem \citep[][D18]{valtonen2011}. This is because the theorem allows us to connect the scaled quadrupole moment $q_2$ and the Kerr parameter $\chi$ of the primary BH by
\begin{equation}
    {q_2}=-q\,\chi^2\,,
    \label{eqn:no-hair}
\end{equation} 
where $q$ should be unity in GR \citep{thorne1980} and therefore testable with present observations.

\section{Observations And Implications of the 2019 Outburst} \label{sec:2019_outburst}

\subsection{Spitzer Observations And Data Reduction} \label{subsec:obs}

\begin{deluxetable*}{lccc}
\tablecaption{
Multi-epoch \texttt{Spitzer} observations in 3.6 $\mu$m (Ch-1) and 4.5 $\mu$m (Ch-2) wavelength bands and the ground-based observations in the optical $R$-band. \label{spitzer-obs-table}}
\tablewidth{0pt}
\tablehead{
\colhead{Epoch (UT)} & \colhead{Ch-1 Flux (mJy)} & \colhead{Ch-2 Flux (mJy) }& \colhead{$R$-band Flux (mJy)} \\
}
\startdata
2019 Feb 25 23:23:06.905 & $17.8\pm0.1$ & $21.8\pm0.1$ & $2.836\pm0.005$ \\
2019 Feb 26 22:02:51.370 & $17.7\pm0.1$ & $21.6\pm0.1$ & $2.825\pm0.005$ \\
2019 Feb 28 01:21:52.252 & $18.2\pm0.1$ & $22.6\pm0.1$ & $2.881\pm0.003$ \\
2019 Mar 01 01:01:21.123 & $17.3\pm0.1$ & $21.2\pm0.1$ & $2.875\pm0.011$ \\
2019 Mar 02 01:39:00.677 & $17.0\pm0.1$ & $20.8\pm0.1$ & $2.825\pm0.013$ \\ 
2019 Jul 31 15:25:33.651 & $26.3\pm0.1$ & $32.3\pm0.1$ & --\\
2019 Aug 01 07:53:36.630 & $26.0\pm0.1$ & $31.7\pm0.1$ & --\\
2019 Aug 01 16:04:46.053 & $26.5\pm0.2$ & $32.0\pm0.1$ & --\\
2019 Aug 02 02:03:48.230 & $25.5\pm0.1$ & $31.0\pm0.1$ & --\\
2019 Aug 02 18:44:48.833 & $24.7\pm0.1$ & $30.0\pm0.1$ & --\\
2019 Aug 03 15:41:47.976 & $25.7\pm0.1$ & $31.3\pm0.1$ & --\\
2019 Aug 04 15:15:32.277 & $24.5\pm0.1$ & $29.9\pm0.1$ & --\\
2019 Aug 05 14:21:42.642 & $23.8\pm0.2$ & $28.9\pm0.1$ & --\\
2019 Aug 06 12:09:24.649 & $23.5\pm0.2$ & $28.9\pm0.1$ & --\\
2019 Aug 07 13:10:27.952 & $23.6\pm0.1$ & $29.0\pm0.1$ & --\\
2019 Aug 08 19:12:35.339 & $24.0\pm0.1$ & $29.2\pm0.1$ & --\\
2019 Aug 09 12:52:32.488 & $24.0\pm0.1$ & $29.6\pm0.1$ & --\\
2019 Aug 10 19:13:55.542 & $24.3\pm0.1$ & $29.8\pm0.1$ & --\\
2019 Aug 13 07:03:26.024 & $23.5\pm0.1$ & $28.8\pm0.1$ & --\\
2019 Aug 16 21:11:01.225 & $23.0\pm0.1$ & $28.5\pm0.1$ & --\\
2019 Aug 20 18:12:49.690 & $24.3\pm0.1$ & $29.7\pm0.1$ & --\\
2019 Aug 22 19:27:51.842 & $23.9\pm0.1$ & $29.3\pm0.1$ & --\\
2019 Aug 25 01:46:53.100 & $25.0\pm0.2$ & $30.6\pm0.1$ & --\\
2019 Aug 28 02:42:37.747 & $22.9\pm0.1$ & $28.4\pm0.2$ & --\\
2019 Sep 02 21:10:17.922 & $22.3\pm0.1$ & $27.7\pm0.1$ & --\\
2019 Sep 04 05:44:12.19  &  --          &   --         &$2.918\pm0.007$ \\
2019 Sep 05 02:44:00.96  &  --          &   --         &$3.179\pm0.092$ \\
2019 Sep 05 02:52:00.48  &  --          &   --         &$3.138\pm0.080$ \\
2019 Sep 06 01:47:09.46  &  --          &   --         &$3.313\pm0.060$ \\
2019 Sep 06 17:43:11.593 & $23.7\pm0.1$ & $29.8\pm0.2$ & --\\
\enddata

\tablecomments{Times are in UT and reflect the start of the observation in Ch-2. The Ch-1 observations started after the Ch-2 observations were done, about 2 minutes and 10 seconds after the start of the ch 2 observations typically. The $R$-band flux observations are not exactly simultaneous with the \texttt{Spitzer} observations but at very close epochs (within 2 hours).}

\end{deluxetable*}

Visibility and scheduling constraints did not permit \texttt{Spitzer} \citep{werner2004} to observe OJ~287 until 2019 July 31, 15 UT, several hours after the predicted time window for the occurrence of the impact flare peak.
Therefore, we focused on the declining part of the expected flare where the radiating bubbles are optically thin in all relevant wavebands \citep{valtonen2019}. 
This part lies between the first brightness peak and the first major minimum, predicted to occur during 2019 August 7. 
The \texttt{Spitzer} scheduling permitted dense monitoring during this critical period. 
Altogether OJ~287 was observed  with \texttt{Spitzer}'s Infrared Array Camera \citep[][]{fazio2004} on 21 epochs between 2019 July 31 and 2019 September 6.
The cadence was approximately 12 hours for the first five epochs, then once per day for the next eight epochs, and thereafter approximately twice a week for the last eight epochs.
Additionally, OJ~287 was monitored on five epochs between 2019 February 25 and 2019 March 2 with daily cadence for normalization purposes with simultaneous optical observations. 
February--March observations permitted us to convert the infrared \texttt{Spitzer}/IRAC channel-1 (3.6 $\mu$m) and channel-2 (4.5 $\mu$m) flux densities, observed during the flare, to equivalent $R$-band flux densities. 
These observations were taken as part of the \texttt{Spitzer} DDT program pid 14206. The observing log and reduced flux densities are given in Table~\ref{spitzer-obs-table}.

All of the observations were taken both in the 3.6 and 4.5 $\mu$m channels (corresponding approximately to the conventional photometric $L$- and $M$-bands) using the 2-second frame time with a 10-position medium-scale dither (typical dither amplitudes of less than an arcminute). The same dither starting point was used in every observation so that OJ~287 landed roughly on the same pixel position in each observation and dither offset.

The corrected basic calibrated data (CBCD) frames were inspected by eye, and remaining artifacts, such as column pulldown, were removed with the {\texttt imclean} tool\footnote{https://irsa.ipac.caltech.edu/data/SPITZER/docs/\\dataanalysistools/tools/contributed/irac/imclean/.}.
Frames with a cosmic detection within a ten-pixel (approximately 12 arcsecond) radius of OJ~287 were not included in the analysis (in general there were zero to one such frames per observation). The centroid of the image of OJ~287 was found with the first moment centroiding method\footnote{https://irsa.ipac.caltech.edu/data/SPITZER/docs/irac/\\calibrationfiles/pixelphase/box\_centroider.pro.}. We performed aperture photometry with the IDL procedure {\it aper} using a source aperture radius of six pixels and a background annulus between 12 and 20 pixel radial distance from the centroid position. We corrected the flux densities with the irac\_aphot\_corr.pro procedure\footnote{https://irsa.ipac.caltech.edu/data/SPITZER/docs/\\dataanalysistools/tools/contributed/irac/iracaphotcorr/.}, for the pixel phase and array location-dependent response functions.
In addition, we performed an aperture correction as tabulated in the IRAC Instrument Handbook\footnote{https://irsa.ipac.caltech.edu/data/SPITZER/docs/\\irac/iracinstrumenthandbook/.}.
For each channel, at each epoch, we finally calculated the mean flux density and the uncertainty from the standard error of the mean, as presented in Table~\ref{spitzer-obs-table}.

\subsection{Extracting the Presence of the Impact Flare and Its Implications} \label{subsec:evidence}

Recall that we predicted the optical $R$-band lightcurve for the 2019 impact flare from the corresponding observations in 2007.
However, our observations of the {\it Eddington flare} are in the two near-infrared \texttt{Spitzer} channels.
Therefore, it is crucial to estimate how the predicted flare lightcurve should look in the \texttt{Spitzer} bands.
In the quiescent state, the infrared--optical wavelength emission comes from synchrotron radiation, and the spectrum follows a power law with a spectral index $\alpha\sim-0.95$ \citep{kidger2018}. 
In contrast, BH impact flares are dominated by bremsstrahlung radiation, which has a nearly flat spectrum in the near-infrared--optical wavelengths superposed on the usual synchrotron emission.
Therefore, we expect that the impact-induced fluxes in the \texttt{Spitzer} bands will be similar to those in the optical bands.
However, the base levels of the fluxes in the  \texttt{Spitzer} and optical wavelengths should be different during such outbursts due to the steep power-law spectrum of the synchrotron background.
Therefore, we subtract the base-level fluxes from the observed \texttt{Spitzer} band fluxes during the outburst to compare with the predicted $R$-band flux curve.
We expect the 2019 impact flare to be coincident in time in the optical and the near-infrared as multi-wavelength observations of the 2015 impact flare show no time delay across the relevant wavebands \citep{valtonen2016, kushwaha2018}.

\begin{figure*}
    \centering
    \includegraphics[scale = 0.7]{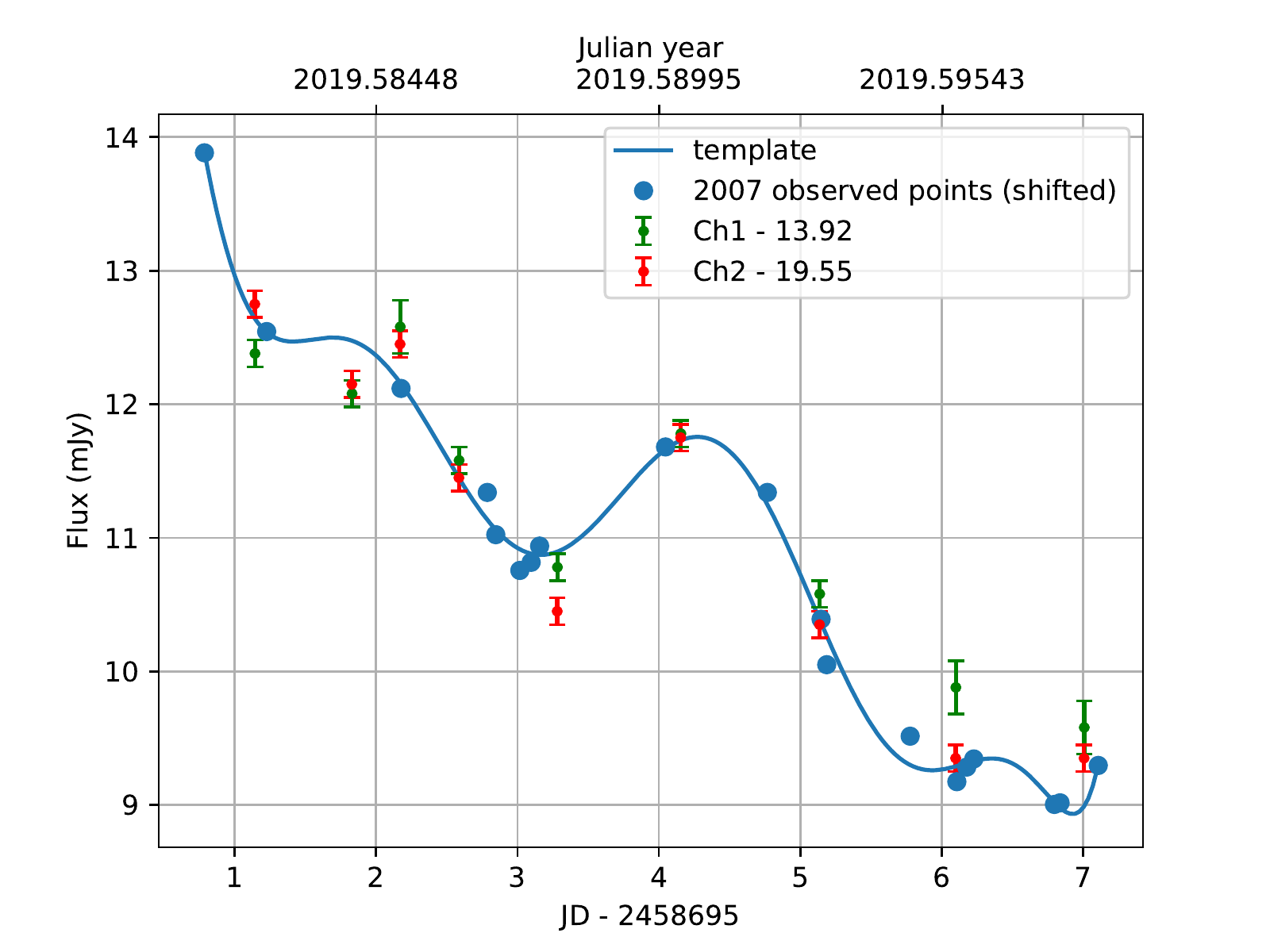}
    \caption{
   Observed \texttt{Spitzer} flux variations of OJ~287 during 2019 July 31 to 2019 August 6 (green and red points with the error bars provide the base-level corrected fluxes in the two near-infrared \texttt{Spitzer} channels). 
   Solid line connects the multi-epoch optical observations (blue filled circles) of the 2007 impact flare, shifted by the predicted $11.8752+(0.06/365.25)\approx11.8754$ years time interval. 
   The temporal shift of our 2007 template does incorporate the fact that the observed flare came 0.06 days later than our prediction. The template is given by a ninth-order polynomial that minimally and smoothly fits the 2007 optical data. An apparent agreement does exist between our prediction and observations.
    }
    \label{fig:flare}
\end{figure*}

We now examine if our observed \texttt{Spitzer} lightcurve does contain the predicted impact flare.
This requires us to create a template of the expected flare, as given in Figure~\ref{fig:flare}, and we  focus on the declining part after the first peak which lasts around seven days.
The template is obtained  by fitting a polynomial to the observed $R$-band lightcurve of the 2007 flare \citep{val_sil11}, shifted forward by $11.8752$ years.
This time shift between the 2007 and 2019 flares, with $\pm 4$ hour uncertainty, was previously found in the orbit solution (D18). 
We introduce the three parameters ${\rm\Delta{t}}$, ${\rm\Delta{F_1}}$ and ${\rm\Delta{F_2}}$ for fitting the \texttt{Spitzer} data with the outburst template. 
The parameters ${\rm\Delta{F_1}}$ and ${\rm\Delta{F_2}}$ are used to correct for the expected base-level differences between $R$-band and \texttt{Spitzer}'s Ch-1 and Ch-2 fluxes, respectively.
The ${\rm\Delta{t}}$ parameter allows us to find the difference between the predicted and actual arrival times of the 2019 outburst. Note that it shifts the time variable in our polynomial fit for the 2007 lightcurve.
We employ only a single ${\rm\Delta{t}}$ parameter for both channels as we expect the impact flare to produce simultaneous flux variations in both channels.
The best-fit values with $1\sigma$ uncertainties read ${\rm\Delta{t}}=-0.06\pm0.05$ days, ${\rm\Delta{F_1}}=13.92\pm0.11$ mJy, and ${\rm\Delta{F_2}}=19.55\pm0.09$ mJy.
This implies that the {\it Eddington flare} arrived $1.4\pm1.2$ hours {\it late} of the predicted epoch but well within the expected time interval.
Therefore, we shift our flux templates forward in time by $0.06$ days to obtain Figure~\ref{fig:flare} where we compare the base-level corrected flux variations in \texttt{Spitzer} channels with the template of 2007.
We also performed  a self-consistency test  by fitting the Ch-1 and Ch-2 fluxes separately and the resulting values of ${\rm\Delta{t}}$'s for the two channels agree with each other within their uncertainties, as required.
Qualitatively,  the predicted lightcurve template for the 2019 impact outburst matches fairly well with the base level corrected fluxes of both  \texttt{Spitzer} channels.
To quantify these similarities, we computed Pearson's~$r$ between the observed  \texttt{Spitzer} data sets and the time-corrected template of Figure~\ref{fig:flare} and found high correlations (Pearson's~$r\sim0.98$).
We repeated this analysis using 20000 random 1 week long OJ~287 lightcurves to rule out the occurrence of high Pearson's~$r$ values due to chance coincidences.

It turns out that the possible template choices introduce $\sim 1$ hour uncertainty in the flare timing.
The template curve of Figure~\ref{fig:flare} should actually be a Gaussian band instead of a single line, since there is always some background noise in the source, and because the 2007 observations have associated error bars. 
Instead of a single template curve there could be any number of alternative ones that fit inside the band of $\pm0.3$ mJy vertical half-width. 
Repeating the above processes with this band instead of a single line widens the error bars in ${\rm\Delta{F}}$ but has no effect on ${\rm\Delta{t}}$ beyond the one-hour additional error.
Further, the radiating bubble that emits bremsstrahlung with a Maxwellian velocity distribution can give the spectral index $\alpha\sim-0.2$, rather than 0.0, the exactly flat spectrum ($\rm{intensity}\sim\nu^\alpha$) if the source has a constant temperature $T$ \citep{karzas1961}. 
However, when we are looking at an expanding bubble, the light travel time is different from different parts of the source, and therefore the spectrum is composed of contributions from different temperatures $T$ within some range $\Delta{T}$.
The intensity depends essentially only on the parameter $u=h\nu/kT$. 
If we employ a reasonable assumption that temperatures $T$ are uniformly distributed over this range, then the intensity is constant over the corresponding  range in frequency $\Delta\nu$ and we get a flat emission spectrum. 
Naturally, details depend on the models of the emitting bubble \citep{pihajoki2016}. It is likely that the spectral index $\alpha$ lies between $-0.2$ and 0.0.
For $\alpha=-0.2$, we get ${\rm\Delta{t}}=0.10\pm0.05$ implying that the flare arrived $2.5\pm1.2$ hours {\it early}.
Thus, considering these uncertainties, the {\it Eddington flare} came within $\sim4$ hours of the predicted time.

\begin{figure*}
    \centering
    \includegraphics[scale = 0.55]{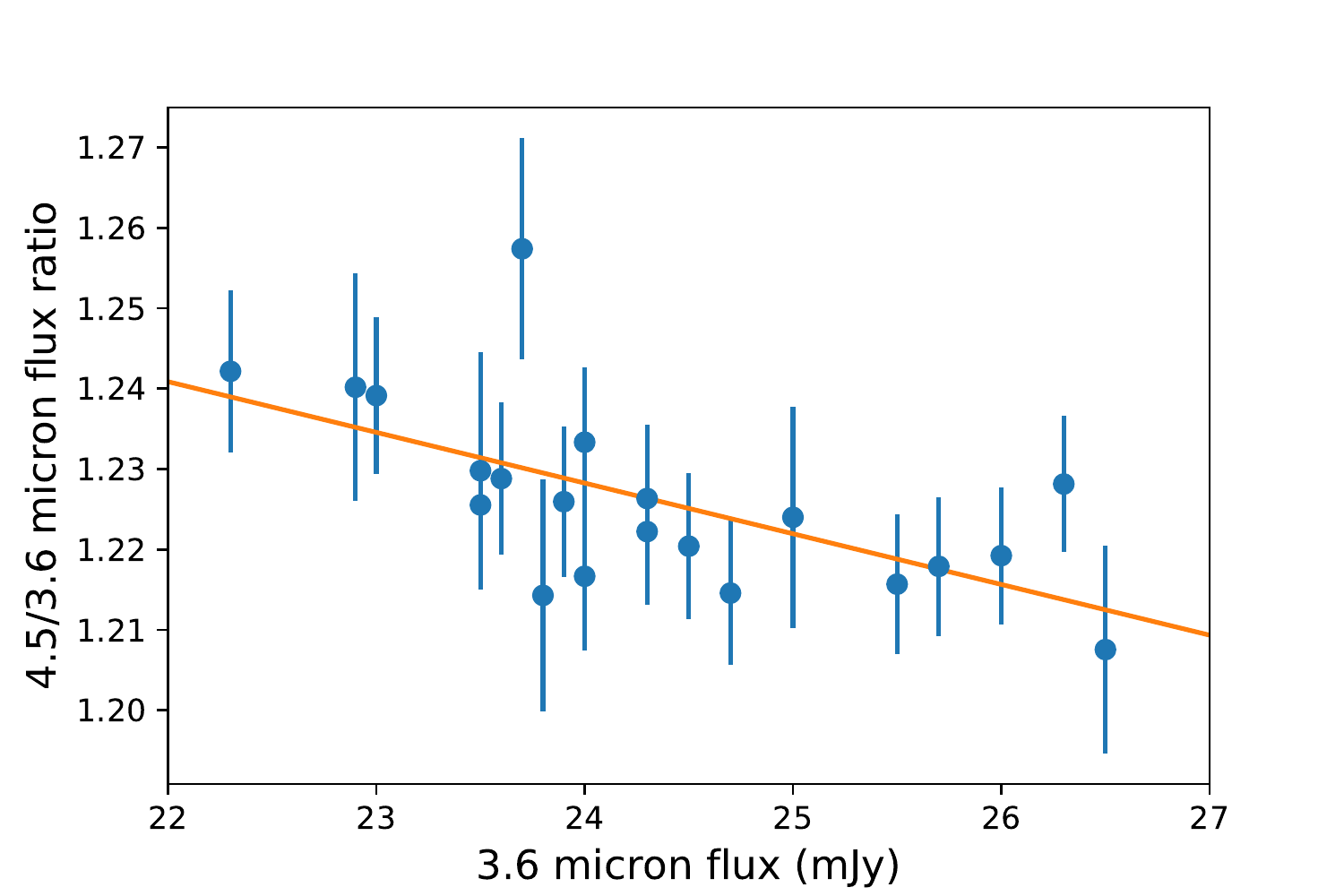}
    \includegraphics[scale = 0.55]{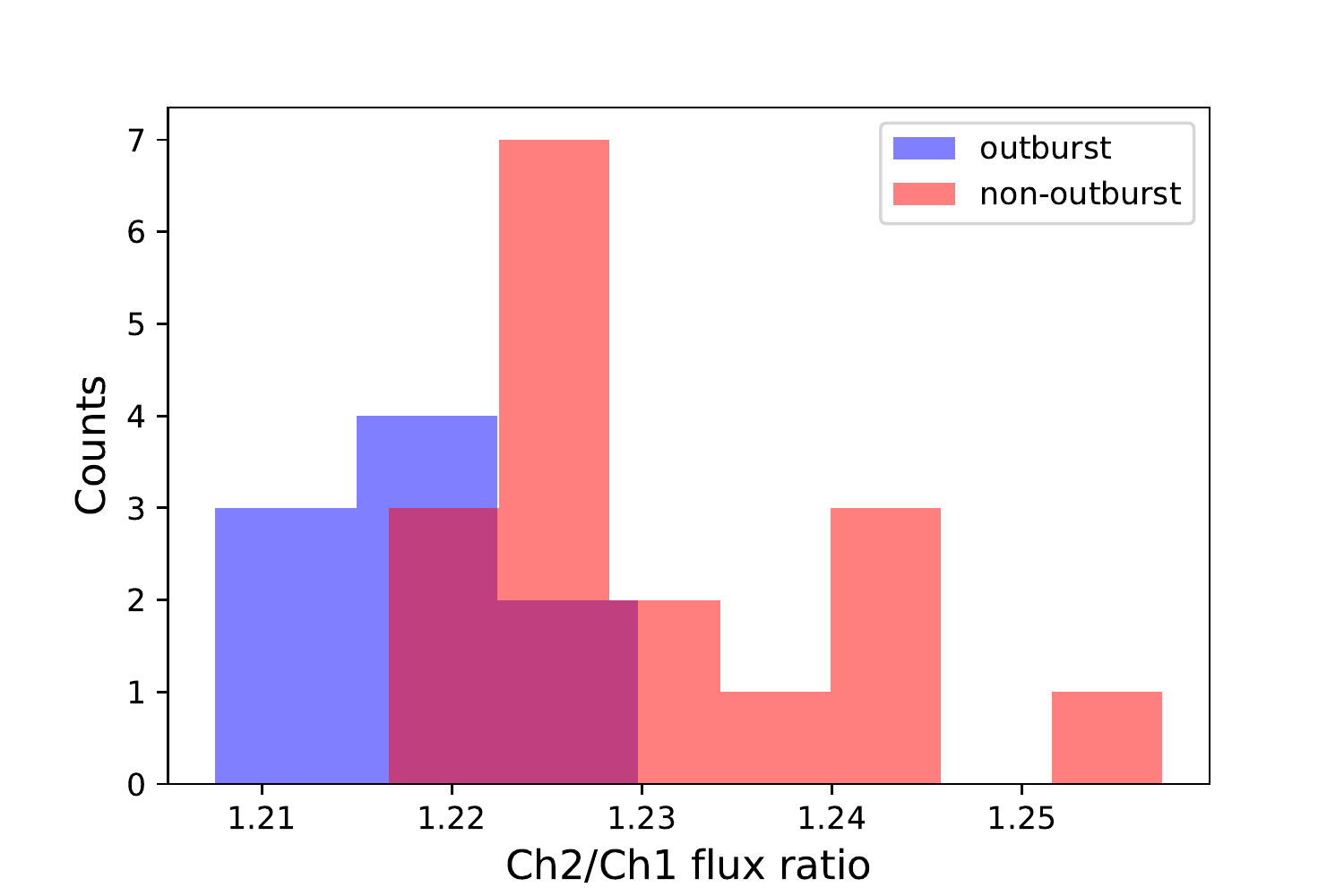}
    \caption{
    Left panel: we display the flux ratio between the $4.5$ and $3.6$ $\mu$m \texttt{Spitzer} channels against the $3.6$ $\mu$m flux, which shows the expected decrease in the ratio when the flux is high.
    Right panel: distribution of the above flux ratio during outburst and non-outburst stretches of data. The bremsstrahlung nature of the flare is responsible for the small flux ratio during flare epochs.
    }
    \label{fig:flux_ratio}
\end{figure*}

\begin{figure*}
    \centering
    \includegraphics[scale = 0.7]{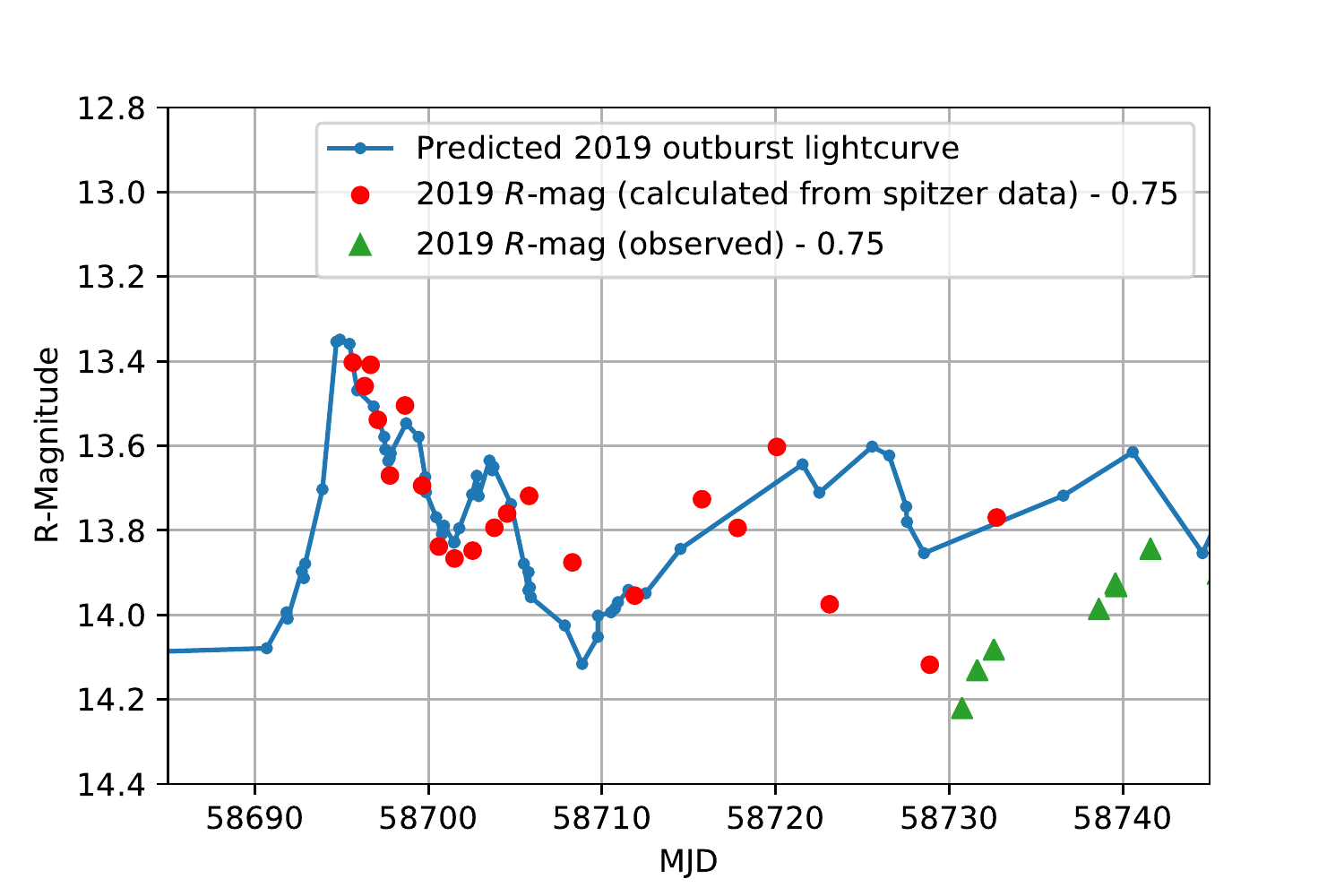}
    \caption{Reconstructed $R$-band lightcurve (red points) and actual $R$-band observation (green triangles) along with the prediction (blue line with dots) (D18).
    The red $R$-band points are constructed from the associated \texttt{Spitzer} fluxes after subtracting the measured 2019 August 16 fluxes as the synchrotron base-level and adding our estimated $R$-band base-level flux of $\sim3.73$~mJy.
    Also, we consider $\alpha=-0.2$ while converting the Spitzer fluxes to $R$-band fluxes.
    }
    \label{fig:R-band_flare}
\end{figure*}

The nearly flat spectrum of the impact flare should cause an overall decrease in the ratio of 4.5 and 3.6 $\mu$m fluxes in the  neighborhood of flare peak in \texttt{Spitzer} data. 
Plots in Figure~\ref{fig:flux_ratio} confirm this expectation.
The flux ratios during the outburst window from 2019 July 31 to August 6 have a significantly different distribution with smaller values compared to their counterparts during the non-outburst phases. 
Further, we got  a Kolmogorov-Smirnov (K-S) statistic of 0.66 with a p-value $=0.0053$ while doing the K-S test between the two set flux ratios. 
At this significance level, the flux ratios during the outburst and non-outburst epochs come from two distinct distributions.

It is also possible to  construct a fiducial $R$-band magnitude flare lightcurve from \texttt{Spitzer} data using the calibration measurements that involved both optical $R$-band and the \texttt{Spitzer} channels during 2019 February 25 -- 2019 March 2.
We find that the Ch-1 flux can be converted to an equivalent optical $R$-band value by dividing the former by $\sim6.2$ and for Ch-2 the factor is $\sim7.6$.
Therefore, using the previously obtained base-level contributions to the infrared flux, the $R$-band base flux during the 2019 impact flare should be $\sim3.73$ mJy.
Thereafter, this $R$-band flux is added to the excess flux above the base-level in the two \texttt{Spitzer} bands.
The resulting two fluxes are averaged at every epoch and converted to R-magnitudes (using Gemini observatory converter).
This is plotted in Figure~\ref{fig:R-band_flare}, together with the actual ground-based $R$-band observations. 
Indeed, the fiducial $R$-band magnitudes join smoothly with the direct $R$-band observations in early September where both \texttt{Spitzer} and optical observations do overlap. 
These plots endorse the similar nature of 2007 and 2019 flares both in their total sizes and general lightcurve shapes.

\section{Discussion} \label{sec:summary}

We presented observational evidence and astrophysical arguments for the occurrence of an impact flare during 2019 July 31 in OJ~287 that was predicted using the BBH central engine model.
These efforts confirm OJ~287 as a source of nano-Hz GWs, which should provide additional motivation for probing the IPTA data sets for GWs from massive BH binaries in general relativistic eccentric orbits \citep{susobhanan2020}.
The present analysis underlines the importance of incorporating the effects of higher-order GW emission in the model.
Interestingly, we would have predicted the flare to occur 1.5 days earlier than it did if we included only the dominant quadrupolar order GW emission in the BBH dynamics.
Observational evidence for the flare arrival within 4 hours of the actual prediction supports the prominent role of including 2PN-accurate GW emission effects while tracking the orbit of the secondary BH.
More importantly, our \texttt{Spitzer} observations constrain the celebrated no-hair theorem by bounding the parameter $q$ in Equation~\ref{eqn:no-hair}. 
The above mentioned timing accuracy corresponds to $q=1.0\pm0.15$ (D18), in agreement with the GR value $q=1.0$, provided  identical impacts generate identical flares, and that the higher-order GW emission is calculated accurately enough.
Such accuracy is possible as our \texttt{Spitzer} observations cover the crucial epoch of fast decline in the flux where the shape of the lightcurve is essentially wavelength independent, which allowed us to tie the variability timescale to the 130 year long record at optical wavelengths. 
These observations are setting the stage for observational campaigns that employ the unprecedented high-resolution imaging capabilities of the Event Horizon Telescope, in combination with the Global Millimeter VLBI Array and the space VLBI mission RadioAstron, to spatially resolve the BBH system in OJ~287.

\acknowledgments We thank Sean Carey for useful discussions on \texttt{Spitzer} data. This work is based in part on observations made with the \texttt{Spitzer Space Telescope}, which is operated by the Jet Propulsion Laboratory, California Institute of Technology under a contract with NASA.
LD and AG acknowledge support of the Department of Atomic Energy, Government of India, under project no. 12-R\&D-TFR-5.02-0200.
SZ acknowledges grant no. NCN 2018/29/B/ST9/01793.

\end{document}